# Fluorescent nanodiamonds for FRET-based monitoring of a single biological nanomotor $F_o F_1$-ATP synthase

M. Börsch*[a], R. Reuter[a], G. Balasubramanian[a], R. Erdmann[b], F. Jelezko[a], J. Wrachtrup[a]

[a]3rd Institute of Physics, University of Stuttgart, Pfaffenwaldring 57, 70550 Stuttgart, Germany;
[b]PicoQuant GmbH, Rudower Chaussee 29, 12489 Berlin, Germany

**ABSTRACT**

Color centers in diamond nanocrystals are a new class of fluorescence markers that attract significant interest due to matchless brightness, photostability and biochemical inertness. Fluorescing diamond nanocrystals containing defects can be used as markers replacing conventional organic dye molecules, quantum dots or autofluorescent proteins. They can be applied for tracking and ultrahigh-resolution localization of the single markers. In addition the spin properties of diamond defects can be utilized for novel magneto-optical imaging (MOI) with nanometer resolution. We develop this technique to unravel the details of the rotary motions and the elastic energy storage mechanism of a single biological nanomotor $F_o F_1$-ATP synthase. $F_o F_1$-ATP synthase is the enzyme that provides the 'chemical energy currency' adenosine triphosphate, ATP, for living cells. The formation of ATP is accomplished by a stepwise internal rotation of subunits within the enzyme. Previously subunit rotation has been monitored by single-molecule fluorescence resonance energy transfer (FRET) and was limited by the photostability of the fluorophores. Fluorescent nanodiamonds advance these FRET measurements to long time scales.

**Keywords:** Nanodiamond, NV center, single-molecule detection, $F_o F_1$-ATP synthase, rotary motor, FRET, FLIM, magneto-optical imaging.

## 1. INTRODUCTION

In search of new photostable fluorescence markers a huge variety of nanoparticles have been investigated over the past years. Probably one of the most exciting materials are diamond nanocrystals. They exist in two forms namely as around 5 nm crystallites from shock wave synthesis or as milled nanoparticles of various size. Both types of nanocrystals are commercially used as e.g. lubricant or polishing agent. Recently nanoparticles have created considerable excitement among material scientists because of their broad range of applications ranging from medical drug delivery to the fortification of polymer materials. This progress has been pushed mainly by considerable advancements in surface chemistry[1]. Especially for biological and medical applications the discovery of brightly luminescent color centers in diamond[2] and diamond nanoparticles have opened a completely new field[3]. Since diamond nanoparticles are of low toxicity[4, 5] the exciting prospect of a universal marker in live cell imaging has appeared. Meanwhile endocytosis and mobility of nanodiamonds in cells has been detected[6]. Specific binding to cellular structures and to DNA as well as proteins was shown.

The Nitrogen-Vacancy (NV) center is a particular type of optically active defect (color centers) which can be excited with 532 nm laser and emits in a broad spectrum centered around 700 nm. It deserves special considerations because of the following:

*First*, the fluorescence of the NV center in diamond is unconditionally stable, that is, it does not bleach or blink even after many months of continuous excitation. As an example of this extreme stability we note that all the data for four recent Science and Nature papers [7-10] published by co-authors of this manuscript were performed on several individual

NVs. These nitrogen-vacancy (NV) centers were studied continuously over a one year period. The microscope setup was eventually automated so that data on individual NVs were taken 24 hours a day. This ability to repeatedly probe a handful of individual color centers on the time scale of years proved critical in achieving the successful results that led to these major advances to the field of quantum computing[11].

*Second*, the NV diamond has the potential to be the smallest nanoparticle probe. All other nanoparticle probes are limited to the range of 10's of nanometers. For quantum dots, even though the dots themselves are only a few nanometers in size, it is necessary to encapsulate them with a thick polymer coating to chemically isolate them from the cell for both optical stability and cell toxicity issues. For colloidal metal nanoparticles, light scattering scales as the inverse sixth power of the size and the detection limit is normally reached for a few 10's of nanometers. Laser-induced heating of metal particles extends the range down to ~10 nm but is limited to this size range due to concerns over high local temperatures. In contrast, NV color centers are stable to within a 1 – 2 nm of the surface of bulk diamond[12], and so the potential exists for NV nanodiamonds of few nanometer diameters.

*Third*, diamond is known for its extreme chemical and biological inertness. For example, diamond nanoparticles are routinely boiled for cleaning in a mixture of sulfuric, nitric, and perchloric acids for hours with no evidence of a chemical reaction. Aside from the toxicity issues of nanoparticles in general, NV diamond nanocrystals are expected to be completely biologically inert. Yet surface functionalization of bulk diamonds with proteins (Horseradish peroxidase[13] and GFP[14]) and DNA[15] has been demonstrated. The native diamond surface can be either hydrophilic, oxygen terminated, or hydrophobic, hydrogen terminated. This gives a very wide range of options for functionalization.

*Finally*, the NV diamond has the potential for extreme spatial resolution down to well below 1 nm. The NV center is unique in that its electron spin is automatically polarized to near 100%, within 100 nanoseconds, when optical excitation is present. Current models for spin polarization invoke the existence of a singlet electronic state whose energy level lies between the ground and excited state triplets (see Fig. 1a). Transitions into this singlet state occur primarily from $m_s = \pm 1$ states, whereas decay from the singlet leads primarily to the $m_s = 0$ ground state. If the remaining optical transitions are spin-preserving, this mechanism should fully polarize the NV center into the $m_s = 0$ ground state. This mechanism also provides the means to optically detect the spin state. Non-resonant excitation (at e.g. 532 nm) excites both the $m_s = 0$ and $m_s = \pm 1$ optical transitions. However, because the intersystem crossing occurs primarily from the $m_s = \pm 1$ excited state, population in $m_s = \pm 1$ ground state undergoes fewer fluorescence cycles before shelving in the singlet state. The $m_s = \pm 1$ states thus fluoresce less than the $m_s = 0$ state, with a difference in initial fluorescence of 20 − 40%. By tuning a microwave field in resonance with its transitions, the spin can also be readily manipulated. These ingredients provide a straightforward means to prepare, manipulate, and measure a single electron spin in a solid at room temperature. The spin polarization can in principle be transferred to nearby electrons[16] and/or nuclei, and this methodology can be potentially applied to bio-molecules to give unprecedented insight into biological macro-molecule dynamics.

Our favorite biological molecule of interest is the nanomotor $F_oF_1$-ATP synthase. This enzyme with 10 nm in diameter and 20 nm in height provides all cells with the "energy currency" adenosine triphosphate, ATP, which is needed for all processes of life[17, 18]. ATP formation is accompanied by an internal rotation of subunits. The $F_1$ part of the enzyme can be removed and works as an ATP-driven rotary motor with 120° steps at high ATP concentrations. Single-molecule videomicroscopy of this motor using large polystyrene beads or µm-long fluorescent biopolymers as a marker of rotation was successfully developed in the last decade[19]. However, viscous drag of these large markers always impaired the high-speed rotation analyses and the angular resolution of the four biochemical processes (ATP binding to one of the three binding sites, ATP hydrolysis, product release of ADP and phosphate). Instead we use two small organic dye molecules which are specifically attached to the rotary and static parts of the enzyme, and measure intramolecular distance changes during rotation by single-molecule fluorescence resonance energy transfer, FRET[20-26]. The single-molecule FRET measurements confirmed the opposite direction of rotation for ATP hydrolysis and synthesis[27, 28], the step size of the rotary subunits in the $F_1$ as well as the $F_o$ motor. We recently applied single-molecule FRET to triangulate the three-dimensional position of the proton-translocating subunit with sub-nanometer resolution[29, 30]. However the general disadvantage of the FRET approach is the short observation time (less than some seconds) which is due to the limited photostability of the dyes.

In contrast to the $F_1$ motor, the directly connected proton-driven $F_o$ motor has a 10-fold symmetry for the bacterial enzyme, and the corresponding step size is 36°. This symmetry mismatch is thought to be of fundamental importance for the "near 100 percent efficiency" of this coupled double motor. The hypothesis of an elastic energy storage by small transient conformational deformations has been proposed[31] and still has to be proven. For the first time, the small but non-bleachable fluorescent nanodiamond marker will enable us to access the angular movement in the rotary parts of a single ATP synthase in a very precise manner and for extended periods of observation times at all ATP / ADP concentrations.

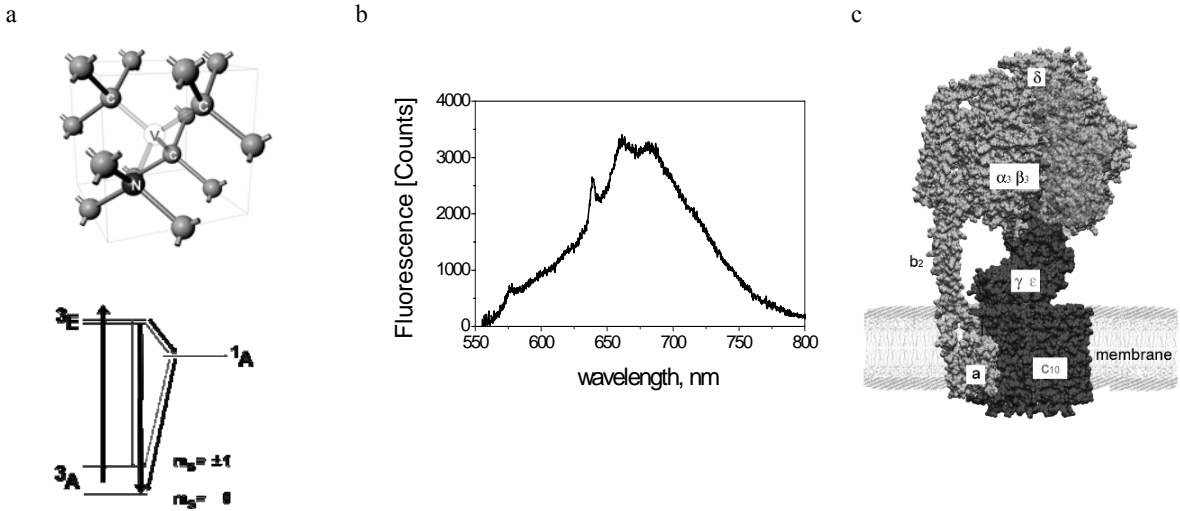

Fig. 1. Left, model of a luminescent Nitrogen-Vacancy (NV) center in the diamond lattice and scheme of electronic state transitions. Bright fluorescence corresponds to the $^3E \rightarrow ^3A$ ($m_s=0$) transition. Middle, fluorescence spectrum of a single NV center in diamond excited with 532 nm. The pronounced peak at 637 nm is the zero-phonon-line. The broadening of the spectrum mainly results from phonon coupling. Right, model of the $F_oF_1$-ATP synthase[32] embedded in the lipid membrane of a vesicle. The $F_1$ motor consists of the non-rotating subunits $\alpha_3\beta_3\delta$ (grey) and the rotating $\gamma$ and $\epsilon$ subunits (black). The $F_o$ motor comprise the static subunits $a$ and $b_2$ (grey) and the rotating ring of ten $c$ subunits (black).

## 2. EXPERIMENTAL PROCEDURES

### 2.1. Fluorescent nanodiamonds by generated defect centers

Fluorescent nanodiamonds can be produced from diamond materials with a high content of nitrogen by irradiating with high energy electrons. The electron beam creates a series of vacancies in the lattice which have to move towards the nitrogen atoms because immediately after irradiation the fluorescent NV centers are not yet formed. This is achieved by annealing the diamonds at 700 °C for about 2 hours. At these temperatures the mobility of the vacancies is high in the diamond. After having created the NV centers the next step is the milling of the diamonds to nanometer-sized particles.

Diamond nanoparticles can be made as small as 1 nanometer using the alternative explosive growth approach. However, those diamonds generally do not have NV centers and, accordingly, are non-fluorescent. A likely reason for this is that the concentration of substitutional unaggregated nitrogen responsible for formation of NV centers in explosion grown diamonds studied so far may be too low. In contrast, type Ib diamonds can have nitrogen concentrations as high as 100 ppm or higher. This corresponds to approximately 5 nitrogen atoms per ten nanometer-sized diamond nanocrystal. Therefore by using a sufficiently high irradiation dose it should be possible in principle to create nanometer-sized diamond crystals with at least one NV per nanocrystal. However, in practice, irradiation causes damage and a too high dose can convert the diamond into graphite. One solution is to irradiate while the diamond is kept at a high temperature. When a diamond is irradiated at temperatures around 700°C damage no longer accumulates due to the mobility of the vacancies at this temperature.

A better solution is to start with ultra-pure diamond and to implant nitrogen. We estimated a NV yield of about ~10 % for this nitrogen implantation approach. Based on a 10% efficiency, a modest nitrogen implant dose of $10^{15}/cm^2$ gives one NV per cubic nanometer which is sufficient to make few-nanometer-sized NV diamond nanoparticles. The additional advantage of starting with ultrapure diamond is that the electron spin linewidth is much narrower in the absence of paramagnetic impurities such as native substitutional nitrogen atoms. This translates to a higher position resolution for MRI imaging.

In addition, many other fluorescent defect centers are known. Ni and Xe implantation generate a whole wealth of defects in diamond with emission at around 800 nm. The NE8 center has very attractive emission properties[33] for cellular application since its emission linewidth is only 2 nm and the emission wavelength guarantees little interference with cellular autofluorescence. Yet it is significantly more challenging to produce. The NE8 defect comprises a substitutional Ni impurity with a complex of 4 surrounding nitrogen atoms. Hence it can only be produced in nitrogen-rich diamond. Initial experiments have been successful in creating NE8 in CVD-grown diamond films. Most of the defects generated show a high fluorescence quantum yield with fluorescent lifetimes in the ns range. Ni and Xe implantation should thus be tested to achieve luminescence in the long wavelength emission range and for possible applications in a FRET assay being the FRET acceptor for the fluorescent NV center in an adjacent nanodiamond.

### 2.2. $F_oF_1$-ATP synthases in liposomes for FRET labeling

The general preparation procedures for $F_oF_1$-ATP synthase from *Escherichia coli* and the exchange of the $F_1$ part has been published[34]. Briefly, the latest $F_oF_1$-ATP synthase mutants used for single-molecule FRET measurements and the FRET measurements of the internal rotor dynamics contained a genetic fusion of the 'enhanced Green Fluorescent Protein' (EGFP) to the C terminus of the non-rotating subunit $a$ [30] of the $F_o$ part and genetically introduced reactive cysteines at the rotating subunits $\gamma$ [35], $\varepsilon$ [28] or $c$ [23], respectively. In previous FRET measurements of $F_oF_1$-ATP synthase the stator subunits were labeled either at the two cysteines in $b_2$, at a specific lysine in $\beta$, or using fluorescent ATP derivatives[36, 37] at catalytic binding site as the second FRET fluorophores (see Fig. 4 below). Because simultaneous labeling of cysteines in two different subunits cannot be done selectively within one step, we preferred to label the subunits of the $F_1$ and $F_o$ parts separately and to recombine the two parts of the enzyme afterwards. Labeling efficiencies for each fluorophore were determined by UV-VIS absorption spectroscopy.

Enzymes were reconstituted in an excess of preformed liposomes with radii of about 100 nm to assure the ratio of a single enzyme per liposome (afterwards we exchanged the unlabeled $F_1$ part with labeled $F_1$). ATP hydrolysis rates were measured for each step as a control. The labeled $F_oF_1$-ATP synthases in liposomes can be stored until use as 2 µl aliquots at -80°C.

### 2.3. Confocal FLIM microscope setup combined with an AFM

The single-molecule detection of NV centers in nanodiamonds was accomplished on a custom-designed scanning confocal microscope combined with an AFM (MFP-3D Asylum Research) which had been modified compared to the previously reported excitation and detection schemes [7, 38-43]. For fluorescence lifetime imaging a new fiber-coupled picosecond pulsed laser at 532 nm (LDH-P-FA-530, up to 80 MHz repetition rate, Picoquant) was available which could be triggered externally. Nitrogen-vacancy defects were excited either with the pulsed laser or with a frequency doubled c.w. Nd:YAG laser (Coherent Compass) focused on to the sample with a high NA objective (Olympus PlanAPO, NA 1.35). Luminescence light was collected by the same objective and filtered from the excitation light using a dichroic beamsplitter (640 DCXR, Chroma) and a long-pass filter (647 LP, Chroma). Photon counting of the filtered light was performed using two avalanche photodiodes (SPQR-14, Perkin-Elmer). We used time-correlated single photon counting electronics for photon registration with 4 ps time resolution for photon anti-bunching measurements, fluorescence lifetime and fluorescence correlation spectroscopy (Picoharp 300 with router, Picoquant). For FLIM, pixel clock, line clock and frame clock signals were generated by the scanning software. Optically detected magnetic resonance measurements were performed using a commercial microwave source (Rhode & Schwarz GmbH, SMIQ 03) amplified by a travelling wave tube amplifier (Hughes 8020H). Commercially available magnetic cantilevers (Team Nanotec) were used for generation of high magnetic field gradients.

## 3. RESULTS AND DISCUSSION

### 3.1. Surface chemistry to attach a nanodiamond to a single $F_oF_1$-ATP synthase

Nanodiamonds with NV centers were produced by milling of irradiated bulk diamond material. Subsequently graphitic and other organic impurities were removed from the diamond surface by boiling in oxidating acids using a mixture of concentrated sulfuric, nitric and perchloric acid. Afterwards the nanodiamonds were neutralized and washed with water. Size selection by consecutive ultracentrifugation steps yielded nanodiamonds of about 20 nm as measured by atomic force microscopy, AFM. Therefore the nanodiamonds had been spin-coated to a glass coverslip using a 1% polyvinylalcohol (PVA) solution. In these nanocrystals, on average two fluorescent defects were found per crystallite. The fluorescence of these defects was stable, showed no intermittences ("blinking") and did not photobleach.

Preventing aggregation of the nanodiamonds in solvents is the prerequisite for the functional surface modifications which are required for specific binding to proteins. The size of the nanodiamonds was controled in aqueous solution measuring the hydrodynamic radius by fluorescence correlation spectroscopy, FCS. Stable hydrosols of nanodiamonds in water were achieved following borane reduction (Fig. 2).

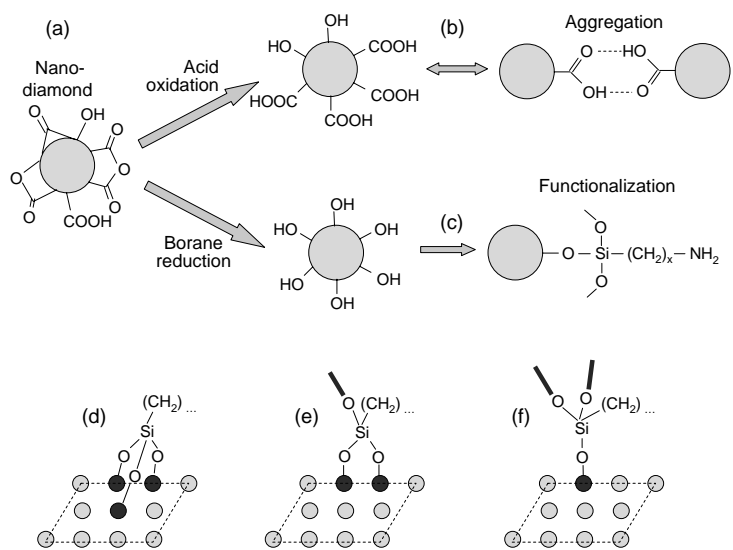

Fig. 2. Steps of chemical surface modifications of nanodiamonds. (**a**) Nanodiamond surface with numerous termination groups. (**b**) Oxidation with strong acids dissolves graphite and organics, leaving only -COOH and -OH terminations. However, -COOH groups form hydrogen bonds leading to aggregation. (**c**) Reduction with borane leaves only -OH surface groups. These are ideal for subsequent silanization with amino-teminated silanes, which are the first step of functionalization. (**d**) Ideal case of all three oxygens linked to the diamond surface. (**e**) Case of two linked oxygens and one dangling linker. (**f**) Case of only one linked oxygen.

The hydoxy groups on the nanodiamond surface are used for silane functionalization. Depending on the OH surface density different binding modes are possible (Fig. 2 d - f). Strongest binding results from linking the silane Si atom via three oxygen atoms to the diamond which leaves a single remaining alkyl chain. If this alkyl chain comprises a functional end group like a reactive NHS ester this can be used for protein binding. However linking the Si atom via two or only one oxygen to the diamond is also advantageous because it allows for cross-linking of the silanes to each other and , thereby, to obtain a quantitative surface coverage.

Until now we have evaluated  a variety of different silanes to modify the diamond surfaces. The functional end groups included an NHS ester for direct but non-specific coupling to the amino groups of proteins. Amino and thiol end groups could be used for attaching spin probes and reactive fluorophores to evaluate FRET. Long polyethyleneglycol linkers terminated with biotin are intended for binding the nanodiamonds to streptavidin-modified proteins and lipids.

### 3.2. Fluorescence lifetime imaging of NV centers in diamond

The fluorescence lifetime of ensembles of NV centers in bulk diamond has been reported previously. Here we measured the fluorescence lifetimes of individual NV centers created in a nanodiamond with a mean size of 10 nm (Fig. 3.) The fluorescence intensity image shows well-separated bright spots which are diffraction limited in size. The distribution of lifetimes from these spots was centered around 17±3 ns (FWHM) which is in accordance with published data.

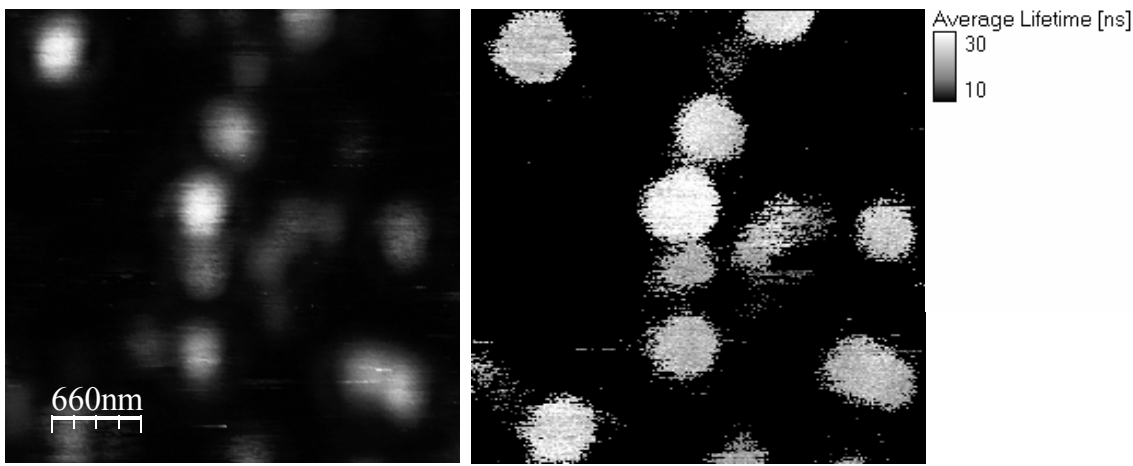

Fig. 3. Fluorescence intensity and lifetime imaging of NV centers in single spin-coated nanodiamonds in PVA (excitation wavelength 532 nm, detection for λ<647 nm). Left, intensity image. Middle, related lifetime image of the NV centers. Right, color code for the intensity image on the left in counts per ms and the lifetime image on the right.

Given the relatively long lifetime of the NV center and its unlimited photostability the potential use of nanodiamonds for FRET-based monitoring of conformational changes in single proteins can be anticipated. Currently we investigate two types of fluorophores being either the FRET donor fluorophore in combination with the NV center serving as the FRET acceptor, or *vice versa* being the FRET acceptor for NV as the FRET donor. Preliminary results indicate the reduction of NV fluorescence lifetimes in the presence of a surface-bound FRET acceptor which was reversed upon photobleaching of the acceptor dye (data not shown).

### 3.3. Towards single-molecule FRET with $F_oF_1$-ATP synthase using a nanodiamond as a label

In the last 10 years we have developed a fluorescence approach to monitor conformational changes and rotary subunit movements in single membrane-embedded $F_oF_1$-ATP synthase from *E. coli*. We apply single-molecule Förster-type fluoresence resonance energy transfer (FRET) using two fluorophores to monitor the rotary motions[44, 45]. The well-known distance dependence of FRET[46, 47] is widely used to map the spacing between two intramolecular[48] or intermolecular[49] positions in the range between 2 and 8 nm. We resolve distance changes between the two fluorophores, one covalently attached to one of the rotating subunits and the second one bound to a non-rotating subunit. Subunit rotation results in a sequence of distance changes between these two fluorophores. The enzmye is incorporated into freely diffusing lipid vesicles with a diameter of 100 - 150 nm. Thereby a single enzyme is investigated at essentially the same conditions as in the biochemical assays to measure ATP synthesis and hydrolysis rates.

In Fig. 4 the different labeling positions for the single-molecule FRET studies are summarized. To monitor γ subunit rotation we labeled γ at a cysteine genetically introduced at residue position 106 [35] and used a non-rotating secondary labeling position at a specific lysine at one of the β subunit[21] (Fig. 4a). Both labels are located on the $F_1$ part. The lysine was labeled selectively with sulfonyl fluoride derivatives of rhodamines at pH 9.0 and 4°C. However, to use the NV center as an alternative modifying the nanodiamond surface with sulfonyl fluoride groups appears to be more complicated using silane chemistry. We favor the non-rotating peripheral dimeric *b* subunits with cysteines[50] to attach a maleimide-modified nanodiamond instead (Fig. 4b). Quantum dots with one maleimide group have been attached to these cysteines[51] and single-molecule FRET to the ε subunit has been demonstrated. Similarly we revealed the 120° stepping of the ε subunit during ATP hydrolysis at 1 mM ATP (Fig. 4c).

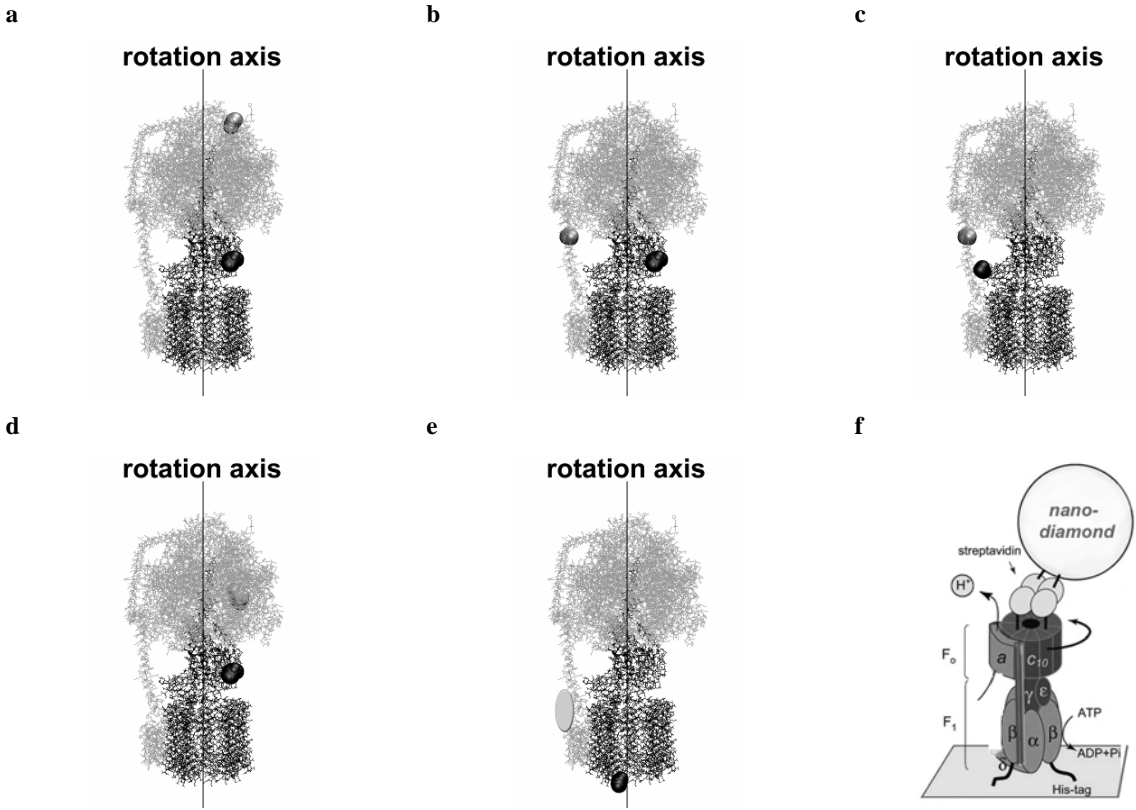

Fig. 4. Fluorophore positions for single-molecule FRET distance changes in $F_oF_1$-ATP synthase. Label positions at the rotating subunits are shown as black dots. Static labeling positions are shopwn as grey dots. a, FRET between labels at lysine 4 (grey dot) on β and γ106 (black dot). b, FRET labels at residues *b*64 (grey) and γ106 (black). c, FRET labels at residues *b*64 (grey) and ε56 (black). d, FRET labels at a bound ATP derivative (grey) and γ106 (black). e, FRET labels at subunit *a* (grey ellipse) and at on one of 10 *c* subunits (black). f, surface-bound, detergent-solubilized $F_oF_1$-ATP synthase with a nanodiamond attached to the rotating *c* subunits.

Once the attachment procedures for the nanodiamonds to the *b* subunits is established one could also use ATP derivatives suitable for FRET to investigate the turnover of ATP at the three catalytic binding sites and study the cooperativity, that is, the increase of rates under muti-site conditions by several orders of magnitude. One way to avoid the photobleaching-limited observation times in single-molecule fluorescence detection is to permanently exchange fluorophores. As has been noticed for many FRET fluorophore pairs the FRET acceptor is more often photobleached before the FRET donor is destroyed. Accordingly we used an fluorescently labeled ATP derivative as an exchangeable FRET acceptor[36] (Fig. 4d). The large fluorophore can partly accommodate in the binding pocket and ATP gets hydrolyzed at slower rates[37]. These single-molecule FRET experiments unraveled more than three FRET levels (or γ subunit orientations, respectively) with respect to one nucleotide binding site indicating the existence of additional stopping positions for γ.

To monitor $c$ ring rotation in a coupled $F_oF_1$-ATP synthase, that is, for the case of an enzyme reconstituted into a lipid vesicle and being capable of ATP synthesis, we fused the autofluorescent protein EGFP to the C terminus of the $a$ subunit[29]. A single EGFP was bright enough for single-molecule detection and for subsequent intramolecular FRET measurements[30]. However, limited photostability and laser power-dependent spectral fluctuations and shifts restricted the use of EGFP as a FRET donor label. The second fluorophore for FRET was attached to a cysteine on one $c$ subunit (see Fig. 4e). Rotation of the $c$ ring during ATP hydrolysis using detergent-solubilized $F_oF_1$-ATP synthase was found to occur in large 120° steps[51, 52]. However, for the proton-driven $c$ ring rotation upon ATP synthesis a step-by-step rotation for each $c$ subunit passing the proton-translocating $a$ subunit was predicted[53]. For ATP synthesis conditions in the presence ADP, phosphate and a pH gradient across the membrane, we clearly identified smaller steps than 120° for the rotating $c$ ring as seen for the $Na^+$ driven ATP synthase[54].

Quantitative FRET distance information can be obtained from the ratio of the measured fluorescence intensities of FRET donor and acceptor with two spectrally distinct detection channels:

$$E_{FRET} = \frac{I_A}{\gamma I_D + I_A} \quad (1)$$

with $I_A$ and $I_D$, fluorescence intensities from the acceptor/donor dye, and γ, correction factor, which consists of the quantum yields of the dyes and the detection efficiency of both detection channels. Alternatively, the lifetime information from the FRET donor dye can be used requiring excitation by a pulsed laser. An increasing FRET efficiency is associated with a reduction of the donor lifetime

$$E_{FRET} = 1 - \frac{\tau_{DA}}{\tau_{D_0}} \quad (2)$$

with $\tau_{DA}$, the fluorescence lifetime of the FRET donor in presence of the acceptor, $\tau_{D_0}$ is the fluorescence lifetime of the donor in the absence of a FRET acceptor or other quenchers in the local environment. Using pulsed excitation allows to control the FRET efficiency calculation based in fluorescence intensities by the simultaneous determination of $E_{FRET}$ based on the single donor fluorophore lifetimes especially for the ultrastable and long-lived NV center in nanodiamonds.

However, to resolve small conformational distortions in $F_oF_1$-ATP synthase very high accuracy of position determinations will be required. The mismatch between the 120° stepping of the γ and ε subunit and the smaller step sizes of the $c$ ring will induce elastic deformations most likely within the rotor[55] parts of $F_oF_1$. This will result in relative distance changes between the ε and the c subunits as well. To monitor these elastic deformations we will use the well-known electronic ground state properties of the NV center in diamond which is a triplet state (see Fig. 1). The fluorescence quantum yield of the NV center depends on the spin state of the system and can be switched by microwaves in the 3 Ghz range. Using external, very steep magnetic gradients as generated by a cantilever tip of an atomic force microscope[7] is the way to achieve sub-nanometer resolution for the position of a rotating nanodiamond bound to $F_oF_1$-ATP synthase (as shown in Fig. 4f). Combining single-molecule FRET and this novel magneto-optical imaging is a promising approach to unravel the fundamental details of the ubiquitous rotary bionanomotor which provides the essential energy currency for life, ATP.

**Acknowledgements**

Financial support by the European Community and the VolkswagenStiftung is gratefully acknowledged.